\documentclass[11pt,a4paper]{article}
\usepackage{jinstpub}

\title{The intelligent \acl{PMT}s for OSIRIS}
\author[a,1]{Feng Gao,}
\author[a]{Tim Kuhlbusch,}
\author[a]{Achim Stahl,}
\author[a,2]{Jochen Steinmann,\note{Corresponding author.}}
\author[a,b]{Cornelius Vollbrecht}
\author[a]{and Christian Wysotzki}
\affiliation[a]{Physics Institute 3B, RWTH Aachen University, Aachen, Germany}
\affiliation[b]{Nuclear Physics Institute IKP-2, Forschungszentrum Jülich GmbH, Jülich, Germany}
\note{Now at Université libre de Bruxelles, Brussels, Belgium.}
\emailAdd{jochen.steinmann@physik.rwth-aachen.de}


\usepackage{amsmath}	
\usepackage{wasysym}	
\usepackage{textcomp}

\usepackage[subrefformat=parens]{subcaption} 

\usepackage{vmargin}	
\usepackage{fancyhdr} 	
\usepackage{longtable}

\usepackage[
    uncertainty-mode = separate             
    ]{siunitx}	                            
    
\usepackage{xparse,xfp}
    
\usepackage[pdftex,dvipsnames]{xcolor}      

\usepackage{multirow}

\usepackage{csquotes}
\usepackage{authblk}
\usepackage{xspace}
\usepackage{mathtools}                      

\usepackage{placeins}                       

\usepackage{todonotes}                      

\usepackage{easyReview}                     

\usepackage[nolist, nohyperlinks]{acronym}  

\usepackage[switch, modulo]{lineno}         

\DeclareSIUnit{\Sps}{Sps}
\DeclareSIUnit{\bar}{bar}
\DeclareSIUnit{\LSB}{LSB}
\DeclareSIUnit{\Samples}{samples}
\DeclareSIUnit{\farad}{F}
\DeclareSIUnit{\photoelectrons}{p.e.}

\newcommand{\Xilinx}{\text{Xilinx}\textsuperscript{\textregistered}\xspace}
\newcommand{\ZYNQ}{\text{Zynq}\textsuperscript{\textregistered}\xspace}

\newcommand{\R}[1]{\ifistoreview \label{#1}\linelabel{#1} \fi}
\newcommand{\lr}[1]{\ifistoreview page~\pageref{#1}, line~\lineref{#1} \fi}

\renewcommand{\replace}[2]{\ifistoreview\remove{#1}~\add{#2}\else #2\fi}
\renewcommand{\remove}[1]{\ifistoreview\alertColor{\st{#1}}\else\fi}

\newcommand{\CUTbaselineN}{18.55}
\newcommand{\CUTbaselineP}{20.70}

\newcommand{\baselineAVGstart}{54}  
\newcommand{\baselineAVGend}{84}    

\newcommand{\IntegrationStart}{92}  
\newcommand{\IntegrationEnd}{122}   

\newcommand{\amplBASELINEmean}{19.59}
\newcommand{\amplBASELINEsigma}{2.407}

\newcommand{\amplCalcSPEheight}{\fpeval{\amplSPEmean - \amplBASELINEmean}}
\newcommand{\amplCalcSPEheightSigma}{\fpeval{sqrt(\amplSPEsigma * \amplSPEsigma - \amplBASELINEsigma * \amplBASELINEsigma)}}

\newcommand{\NPEinHG}{\fpeval{(255 - \amplBASELINEmean) / (\amplSPEmean - \amplBASELINEmean)}}
\newcommand{\NPEinHGsigma}{ \fpeval{ sqrt(\termC + \termF)} }

\newcommand{\termA}{\fpeval{\amplSPEmean - \amplBASELINEmean} }
\newcommand{\termB}{\fpeval{ (255 - \amplBASELINEmean) / (\termA * \termA)}}
\newcommand{\termC}{\fpeval{ \termB * \termB * \amplSPEsigma * \amplSPEsigma }}
\newcommand{\termD}{\fpeval{ \amplBASELINEmean - \amplSPEmean}}
\newcommand{\termE}{\fpeval{ (\amplSPEmean - 255) / ( \termD * \termD ) }}
\newcommand{\termF}{\fpeval{ \termE * \termE * \amplBASELINEsigma * \amplBASELINEsigma}}

\newcommand{\amplSPEmean}{51.8}
\newcommand{\amplSPEsigma}{6.0}

\newcommand{\sigmaNoise}{28.05} 
\newcommand{\sigmaNoiseUncert}{0.03} 

\newcommand{\sigmaSPE}{116.6}   
\newcommand{\sigmaSPEUncert}{1.3} 

\newcommand{\gainSPE}{415.2}    
\newcommand{\gainSPEUncert}{1.2} 

\newcommand{\PeakToValley}{3.55}

\newcommand{\chargeSNR}{\fpeval{\gainSPE / \sigmaNoise }}
\newcommand{\chargeSNRUncert}{\fpeval{\chargeSNR * sqrt((\gainSPEUncert/\gainSPE)^2 + (\sigmaNoiseUncert / \sigmaNoise)^2)}}

\newcommand{\chargeNoiseWidth}{\fpeval{ \sigmaNoise / \gainSPE}}

\newcommand{\chargeResolution}{\fpeval{100*\sigmaSPE / \gainSPE} }
\newcommand{\chargeResolutionUncert}{\fpeval{\chargeResolution * sqrt( (\sigmaSPEUncert/\sigmaSPE)^2 + (\gainSPEUncert/\gainSPE)^2)}}

\abstract{\add{The dimensions of liquid scintillator neutrino detectors have grown over the years. Thus, the analog cable length required for the readout increases.}
\remove{With larger and larger detectors --  especially in the regime of neutrino experiments -- the analog cable length required for the readout increases.}\R{A8}
In consequence, the signal loss increases as well. With the intelligent \acl{PMT} this problem can be solved. The digitiser is moved as close as possible to the \acl{PMT}. This concept allows nearly lossless digitisation of the analog signal.
Furthermore, a computing unit next to the \acl{ADC} enables implementation of on-the-fly data processing and control algorithms. 
This paper presents the detailed concept and the performance of the intelligent \acl{PMT}.
}

\begin{document}
\setreviewsoff

\ifistoreview\linenumbers\fi

\begin{acronym}
\acrodef{JUNO}[JUNO]{\textbf{J}iangmen \textbf{U}nderground \textbf{N}eutrino \textbf{O}bservatory}
\acrodef{OSIRIS}[OSIRIS]{\textbf{O}nline \textbf{S}cintillator \textbf{I}ntrinsic \textbf{R}adioactivity \textbf{I}nvestigation \textbf{S}ystem}
\acrodef{Serappis}[Serappis]{\textbf{SE}arch for \textbf{RA}re \textbf{PP}-neutrinos \textbf{I}n \textbf{S}cintillator}
\acrodef{LS}[LS]{liquid scintillator}

\acrodef{DAQ}[DAQ]{data acquisition}

\acrodef{CD}[CD]{central detector}

\acrodef{PMT}[PMT]{Photomultiplier Tube}
\acrodef{TTS}[TTS]{transition time spread}

\acrodef{SMT}[SMT]{Surface Mount Technology}

\acrodef{iPMT}[iPMT]{intelligent \acs{PMT}}

\acrodef{SCCU}[SCCU]{Slow Control and Configuration Unit}
\acrodef{JTAG}[JTAG]{Joint Test Action Group}
\acrodef{I2C}[I2C]{Inter-Integrated Circuit}
\acrodef{UART}[UART]{Universal Asynchronous Receiver Transmitter}
\acrodef{XVC}[XVC]{Xilinx Virtual Cable}
\acrodef{UDP}[UDP]{User Datagram Protocol}
\acrodef{TELNET}[Telnet]{Teletype Network}

\acrodef{SB}[SB]{Surface Board}
\acrodef{ROB}[ROB]{Readout Board}

\acrodef{POE}[PoE]{Power Over Ethernet}
\acrodef{IC}[IC]{integrated circuit}
\acrodef{ASIC}[ASIC]{application-specific integrated circuit}
\acrodef{CMOS}[CMOS]{complementary metal–oxide–semiconductor}
\acrodef{RAM}[RAM]{Random-Access Memory}
\acrodef{CDR}[CDR]{clock data recovery}

\acrodef{HG}[HG]{high gain}
\acrodef{MG}[MG]{mid gain}
\acrodef{LG}[LG]{low gain}

\acrodef{ADC}[ADC]{analog-to-digital converter}
\acrodef{TIA}[TIA]{transimpedance amplifier}
\acrodef{LVDS}[LVDS]{low-voltage differential signaling}
\acrodef{PLL}[PLL]{phase-locked loop}
\acrodef{SoC}[SoC]{System on Chip}

\acrodef{PE}[p.e.]{photo electron equivalent}
\acrodefplural{PE}[p.es.]{photo electron equivalents}

\acrodef{FPGA}[FPGA]{Field Programmable Gate Array}
\acrodef{PRBS}[PRBS]{pseudorandom bitstream}

\acrodef{EPICS}[EPICS]{Experimental Physics and Industrial Control System}
\acrodef{IOC}[IOC]{Input / Output Controller}

\acrodef{HV}[HV]{high voltage}
\acrodef{PMMA}[PMMA]{Poly(methyl methacrylate)}
\acrodef{HDPE}[HDPE]{high-density polyethylene}

\acrodef{FWHM}[FWHM]{full width at half maximum}
\acrodef{SNR}[SNR]{signal to noise ratio}

\end{acronym}

\maketitle

\section{Introduction}
The \ac{JUNO} experiment is a dedicated neutrino detector currently under construction in southern China~\cite{JUNO_YellowBook}. The detection mechanism is based on neutrino interactions with \ac{LS}. A highly radiopure \ac{LS} is of foremost importance for the success of \ac{JUNO}~\cite{JUNO_radioActivity2021}.
Therefore, the dedicated \ac{OSIRIS} experiment monitors the radiopurity of the scintillator during filling of \ac{JUNO}.

\ac{OSIRIS} uses a novel \ac{PMT} integration scheme -- called \ac{iPMT} -- to detect light from radioactive decays.
The concept, realisation and performance of the \ac{iPMT} system is presented in this paper.

\R{A322_fix}Due to unforeseen problems and the necessary repairs of the \acp{iPMT}, they are not installed in the first phase of \ac{OSIRIS}. They will be used in the future upgrade of \ac{OSIRIS} focusing on the search for solar $pp$ neutrinos \cite{SERAPPISconcept}.
\section{\acs{OSIRIS}}
\label{sec:osiris}
The \ac{OSIRIS} experiment is a facility to monitor the internal radiopurity of the scintillator used in the \ac{JUNO} experiment during filling. For this purpose, a fraction of the \ac{LS} is fed continuously into \ac{OSIRIS}~\cite{OSIRISDesign}.
The commissioning of the \ac{LS} purification systems of \ac{JUNO} is also monitored by \ac{OSIRIS}.

\begin{figure}[htb]
    \centering
    \includegraphics[width=0.6\textwidth]{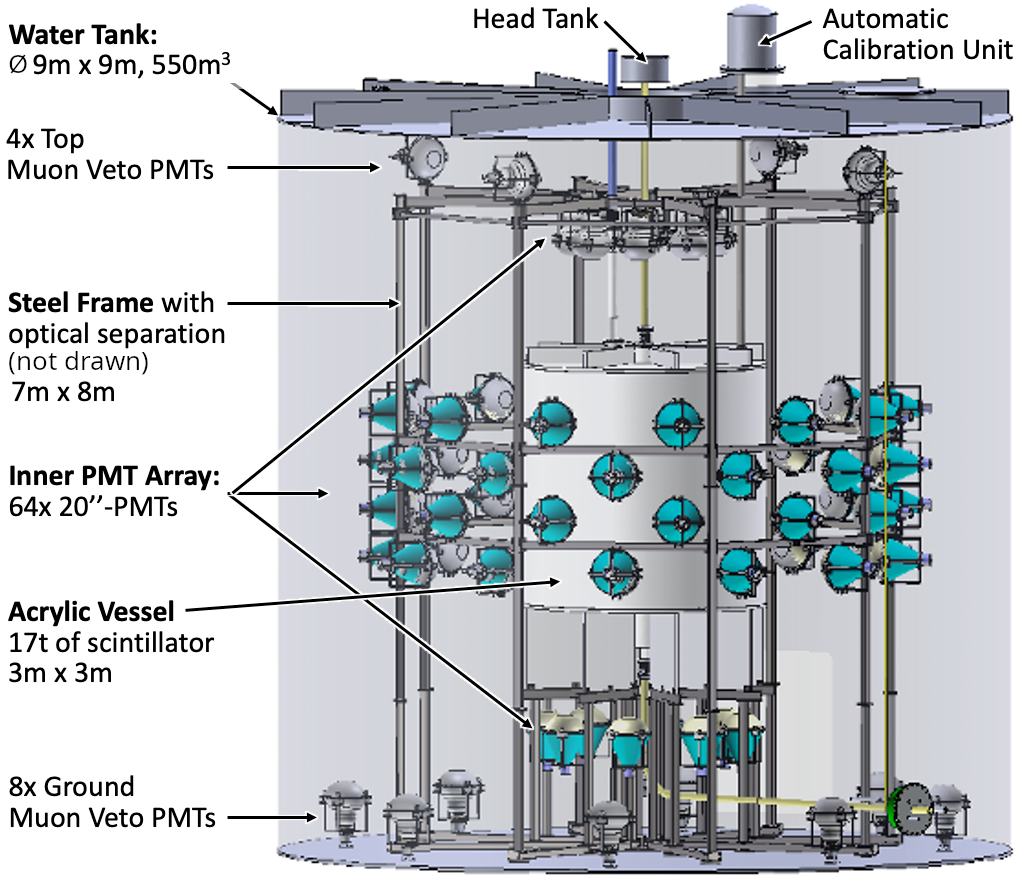}
    \caption{Layout of the OSIRIS detector~\cite{OSIRISDesign}.}
    \label{fig:osiris_layout}
\end{figure}

The detector is separated into the inner detector, which consists of the \ac{LS} contained in the acrylic vessel and the outer detector, that uses the shielding water as a Cherenkov muon veto. The light produced in both detector parts is detected by \acp{PMT}. 
The outer detector is instrumented by \qty{12}{\acsp{PMT}}. Further \qty{64}{\acsp{PMT}} are pointing towards the center of the acrylic vessel (cf. Figure~\ref{fig:osiris_layout}).

\todo[inline, disable]{Extend a little bit here? E.g. there is an acrylic vessel mentioned in the last sentence which is not introduced}

\todo[inline,disable]{Include Fig 2 of the OSIRIS paper referenced}
\section{Concept of the \acl{iPMT}}
\label{sec:concept}
For the readout of large liquid scintillator detectors, long analog cables have been used in the past. For example, Borexino uses a cable length of \qty{57}{\meter} for all \acp{PMT} \cite[p. 12]{TheBorexinoDetector}. A comparable cable length is used in the SuperKamiokande detector \cite{SuperKelectronicsCalibration}. If the size of the detector gets larger, the length of those cables increases. Thus, the attenuation along these cables increases as well. 

Assuming an {RG58} cable with typical attenuation at \qty{100}{\mega\hertz} of \mbox{\qty{12}{\dB}/\qty{100}{\meter}}. At a length of \qty{60}{\meter} the output voltage is reduced to \qty{43.6}{\percent} at \qty{100}{\mega\hertz}.
The attenuation depends on the frequency. Typically, the high frequencies are attenuated more, resulting in a shaping of the signal. Recovery of the original signal is only possible with enormous \replace{afford}{effort}. 
The attenuated frequencies have to be amplified to compensate the attenuation. 
In case the amplification and the attenuation do not match, the original signal is not recovered.

The idea behind the concept of the \acp{iPMT} is to move digitiser and readout as close as possible to the \ac{PMT}. By this, the analog signal path can be reduced to a few centimeters. With the digital output, the cable length no longer has an impact on the quality of the data. The length of the digital cable is limited as well, but the digital signals can be refreshed without loss of information.

\begin{figure}[htb]
    \centering
    \includegraphics[width=0.9\textwidth]{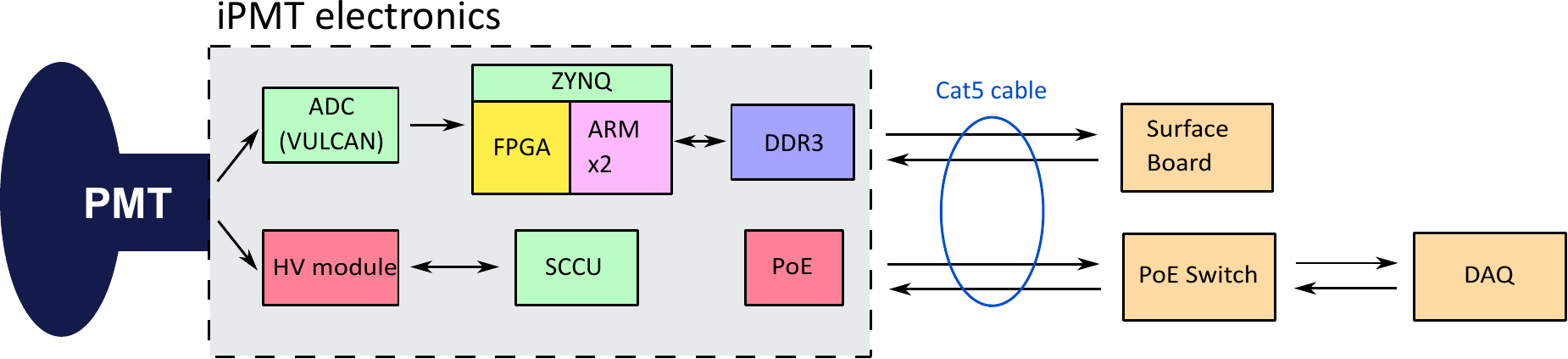}
    \caption{Overview of \ac{iPMT} electronics.}
    \label{fig:ipmt_electronics}
\end{figure}

In order to handle the digitised data, a signal processing unit is inserted at the back of each single \ac{PMT} as illustrated in Figure~\ref{fig:ipmt_electronics}. The processing unit packs the digitised data into waveforms, which are transmitted digitally via Ethernet. In addition, this unit provides resources for data reduction or reconstruction. Also high level algorithms, e.g. a stabilisation of the gain of the \ac{PMT}, can be implemented on the processing unit.

\todo[inline, disable]{Do we want to have an overview of the iPMT electronics (this graph needs obviously some extension)}

The waveforms are transmitted on a self-triggered basis. If the processing unit detects a pulse in the waveform \add{by e.g. a simple threshold discriminator or more advanced pulse detection techniques}\R{A80}, this waveform packet is sent to the \ac{DAQ}. Inside the \ac{DAQ} software, a global trigger and event building is implemented and noise hits are discarded.

The whole \add{readout}\R{A83} concept was initially developed for usage in the \ac{JUNO} central detector \cite{Bellato_2021}. After it became clear that the concept is not going to be used for this detector, the whole concept was iterated again. This second iteration is presented in this paper.

\graphicspath{{HV/img}{BASE/img}{ROB/img}{SurfaceBoard/img}{PMT/img}{performance/img}{concept/img}}
\section{Implementation of the \acl{iPMT}}
\begin{figure}[htb]
    \centering
    \includegraphics[width=0.8\textwidth]{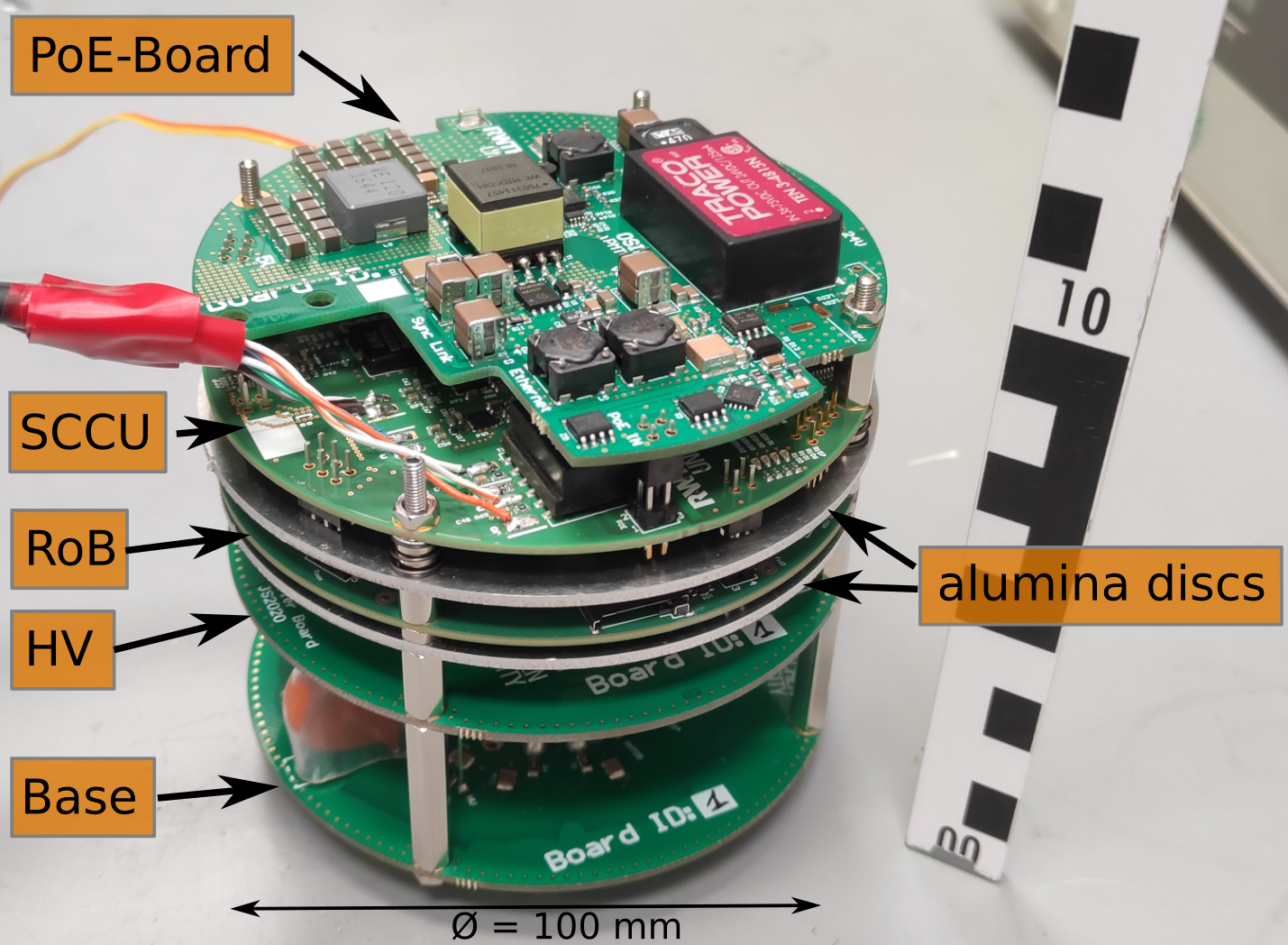}
    \caption{Image of an \ac{iPMT} electronics stack. Alumina discs are placed above and below the \ac{ROB} for a better heat transfer.}
    \label{fig:img_ipmt_electronics}
\end{figure}

In the following sections, each component of the \ac{iPMT} -- from the \ac{PMT} to the \ac{DAQ} -- is described in more details.
A complete assembly of an \ac{iPMT} electronics stack is depicted in Figure~\ref{fig:img_ipmt_electronics}.

\subsection{\aclp{PMT}}
For \ac{OSIRIS}, 76 dynode photomultipliers of the type R15343 from Hamamatsu Photonics were prepared. \remove{Due to unforeseen problems with the {HV} unit, they need to be repaired and will be installed in a future upgrade of {OSIRIS} }.\R{A90}

The PMTs are comparable to the Hamamatsu R12860HQ regarding the dimensions and performance.
\begin{figure}[htbp]
    \centering
    \subcaptionbox{\label{fig:pmt_dist_drc}}{\includegraphics[width=0.49\textwidth]{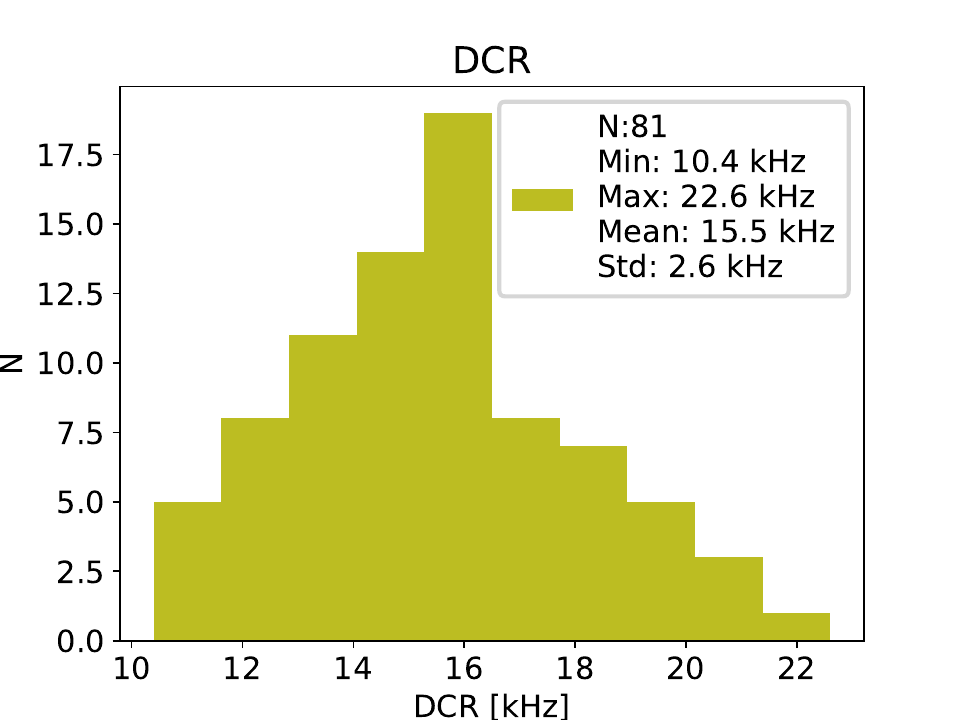}}
    \subcaptionbox{\label{fig:pmt_dist_tts}}{\includegraphics[width=0.49\textwidth]{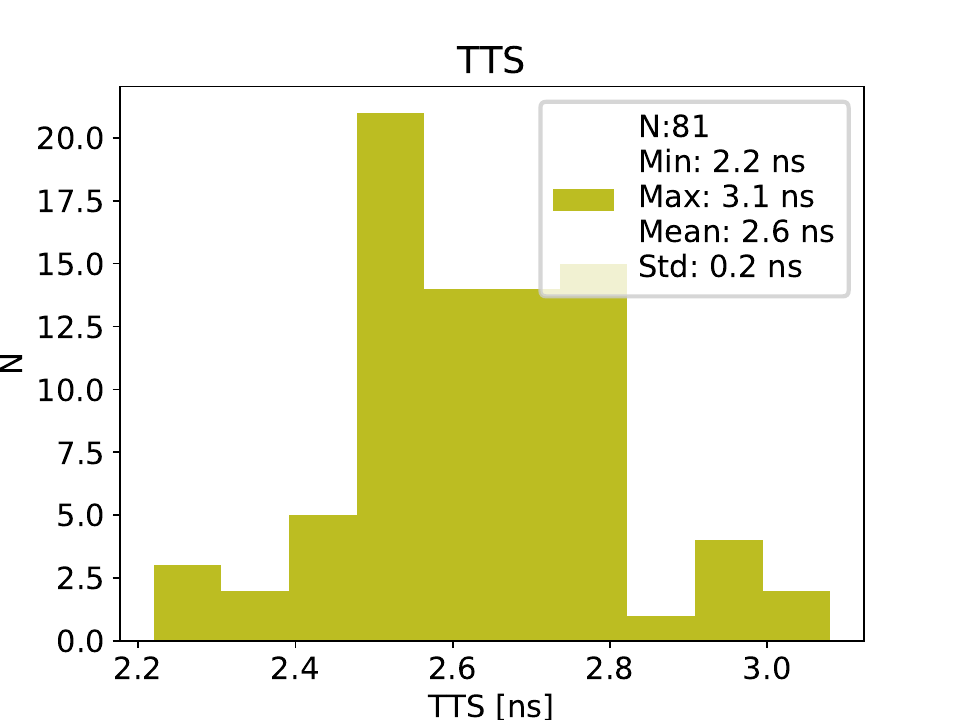}}
    \caption{Distribution of \ac{PMT} parameters as measured by Hamamatsu Photonics: \subref{fig:pmt_dist_drc} dark count rate,   \subref{fig:pmt_dist_tts} transit time spread (\ac{FWHM}).}\label{fig:pmt_dist}
\end{figure}
\ac{OSIRIS}'s \acp{PMT}  can be separated into two groups, the \acp{PMT} looking inwards towards the acrylic vessel, and those, which monitor the water pool and serve as a muon veto.
%
%
12~\acp{PMT} are selected for the muon veto. The selection is based on the data given by Hamamatsu (cf. Figure~\ref{fig:pmt_dist}). There are eight \acp{PMT} with \ac{TTS} of \qty{2.8}{\nano\second} or more. These \acp{PMT} are selected for the veto. The remaining four \acp{PMT} are the ones, which have the highest dark count rate.

\subsection{Voltage Divider (Base)}
The voltage divider follows the reference design by Hamamatsu with a few modifications. For a safe operation, the clearance between the \ac{HV} and the ground traces is adapted according to the voltage. This gives the maximum ground trace thickness. Except for one \qty{10}{\nano\farad}/\qty{6}{\kilo\volt} capacitor all components are \ac{SMT} types. The base itself gets soldered to the connector of the \ac{PMT}. 


\subsection{High Voltage Board}

The high voltage board hosts a high voltage module \cite{Bellato_2021}. The decoupling of the signal from the \ac{HV} is done on this board as well. In order to protect the subsequent components, two gas discharge tubes are added to this board. They limit the voltage to a maximum of \qty{70}{\volt}. To prevent charge-up of the secondary side of the decoupling capacitor, a high ohmic discharge resistor is installed. 
The influence on the signal by the discharge resistor and tubes is negligible.


\subsection{\acl{ROB}}
\begin{figure}[hbtp]
\centering
\includegraphics[width=0.7\textwidth]{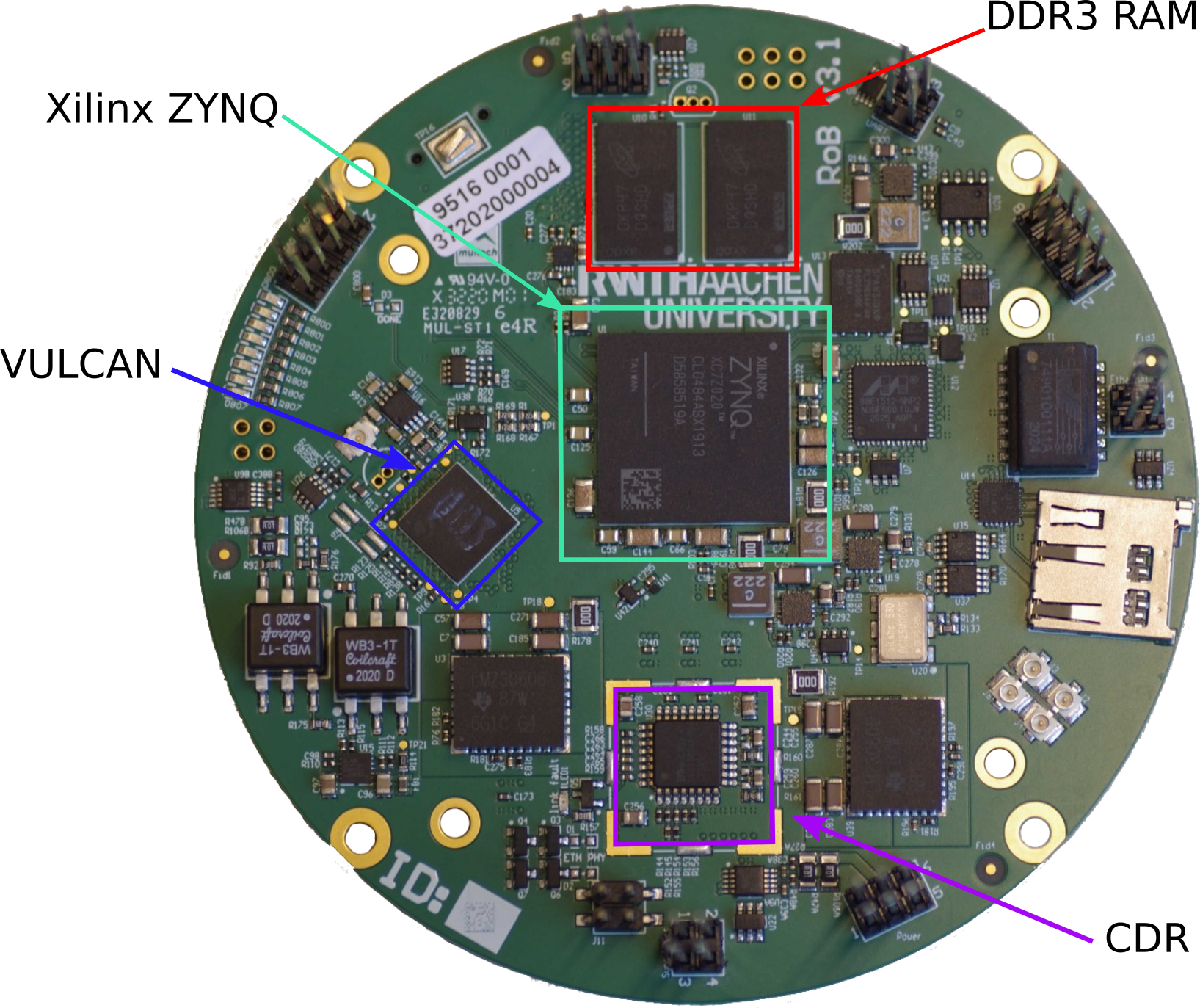}
\caption{Image of the ROB with important \acp{IC} highlighted.}
\label{fig:rob_board}
\end{figure}

The \ac{ROB} is the main board in the readout chain (cf.  Figure~\ref{fig:rob_board}). It digitizes the signal, provides the timestamps of the waveform packets, and contains the digital processing unit, a \Xilinx~\ZYNQ\footnote{FPGA (Artix-7) and a Dual-Core ARM Cortex-A9 MPCore.}.
For dataprocessing, \qty{1}{\giga\byte} of \ac{RAM} is connected~\cite{WysotzkiPhD}.

All \acp{iPMT} are synchronised by a dedicated synchronous link running at \qty{125}{\mega\bit\per\second}. This link embeds a data stream and a clock via Manchester Coding~\cite[142]{Stallings2007}. On the \ac{ROB}, this stream is processed by a \ac{CDR} \ac{IC} (Microchip SY87700AL) and split into a data and clock signal again.

\subsubsection*{VULCAN}
\emph{VULCAN}~\cite{MuralidharanPhd, PhDParkalian} is a dedicated \ac{PMT} frontend and digitization chip developed by the ZEA-2 of the Forschungszentrum Jülich. 
It can be configured via the \ac{JTAG} interface. VULCAN hosts a dedicated \ac{PLL}, which reduces the phase noise \add{of the internal clocks}~\cite{PhDParkalian}\R{A123}. \replace{Although VULCAN is designed to run at a speed of {1}{GSps}, in {OSIRIS} it is used at half the speed.}{Inside \ac{OSIRIS} VULCAN is used with a sampling rate of \qty{500}{\mega\Sps}.} \remove{This reduces the noise of the digitiser.}

VULCAN features three identical receivers. They are configured to cover three ranges: \ac{HG}, \ac{MG} and \ac{LG}. Using these three ranges, VULCAN provides a dynamic range up to \qty{1000}{\photoelectrons} while maintaining a high resolution at lower charges. The \ac{HG} and \ac{MG} channel are using a \ac{TIA} -- with the \ac{HG} channel is configured with a larger amplification. The \ac{LG} channel measures the voltage across the \ac{TIA} input.

Each receiver channel serves \remove{four {6}{bit} {ADC}, which are concatenated and form a single} \add{one} \qty{8}{\bit} \ac{ADC}. All three channels are digitised in parallel and the \replace{processor}{data processing logic}\R{A132} inside VULCAN selects the unsaturated samples with the highest resolution. These samples are forwarded to the \ac{FPGA} via a \ac{LVDS} connection. 

\subsection{\acl{SCCU}}
The \ac{SCCU} provides access to all slow monitoring sensors as well as the low level and debugging access to VULCAN and the \ac{FPGA}.

This board hosts \replace{an}{a}\R{A139} \emph{STM32F417} microcontroller, which provides though its firmware access to \ac{JTAG}, \ac{I2C}, and \ac{UART} via Ethernet. This board hosts a four port Ethernet switch joining the datastreams from the \ac{ROB} and the \ac{SCCU} \add{itself}\R{A142}. One port is the uplink to the \ac{SB}, one connects to the \ac{ROB}, the third one is for the microcontroller, while the fourth one is not used. The access to \ac{JTAG} is implemented using the \ac{XVC} \cite{XVC}. This allows usage of the full \ac{FPGA} tool chain. Via a custom \add{defined protocol over} \ac{UDP}\R{A145}, one can access two \ac{I2C} busses, which are connected to many sensors, which monitor the systems behaviour. The three \ac{UART} connections are accessible via \ac{TELNET} protocol. For each connection, the baudrate as well as the new line character can be configured remotely. The configuration is stored in the internal EEPROM of the microcontroller. Two of the \acp{UART} are routed to the \ac{FPGA} and the third one is converted into \emph{RS485} and connected to the HV module.
\subsection{\acl{POE}}
The whole \ac{iPMT} is powered by \ac{POE} using \ac{POE} Plus (\mbox{IEEE~802.3at-2009}) Class 4. This allows the device to consume up to \qty{25.5}{\watt}. Eventhough this limit is never reached by the \ac{iPMT}, it provides some margin for future extensions of the \ac{FPGA} functionality.

The \ac{POE} signal is rectified using an ideal diode bridge and then fed into the \ac{POE} power delivery protocol handling \ac{IC}. The \qty{48}{\volt} rail is connected to two DC/DC converters. One provides an isolated \qty{24}{\volt} output for the \ac{HV} module. The other one is an isolated \qty{5}{\volt} regulator supplying the \ac{SCCU} and the \ac{ROB}. On these boards, the \qty{5}{\volt} is regulated down to the voltages needed on these boards. On the \ac{POE} board, a temperature monitoring for the transformer of the \qty{5}{\volt} regulator is implemented. The supply voltage and current of the HV module is measured, too. These sensors allow continuous monitoring of the operational conditions.

\subsection{Cable}
The connection between the \ac{iPMT} and the \acf{SB} is a single CAT5 cable with four twisted wire-pairs.
Two pairs are used for the \qty{100}{\mega\bit\per\second} Ethernet with embedded \ac{POE}. The third pair hosts the synchronous connection from the SB to the iPMT, whereas the fourth pair is used for synchronous transmission from the iPMT to the SB.

The whole CAT5 cable is enwrapped in a metal mesh shielding and \replace{an}{a} \ac{HDPE} coat. This outer mantle is compatible with the ultra-pure water in the detector tank. The \ac{HDPE} coat is filled with a water blocking powder. In case of damage, the powder prevents propagation of water along the cable.

For \ac{OSIRIS}, a maximum cable length of \qty{25}{\meter} is required~\cite{OSIRISDesign}. During installation of the \acp{iPMT}, this cable can be cut to the right length. The technical limit is given by the Ethernet requirements. Typically this limit is \qty{90}{\meter}. During the production tests, the cable has been elongated by additional \qty{20}{\meter}. No communication problem could be observed with this configuration.
\subsection{Potting \& Holder}
\label{sec:potting_holder}
The \acp{iPMT} will be submerged in a pool of ultra-pure water. The electronics need to be encapsulated to protect it from contact with water and to keep the water clean. 
The maximum water depth in \ac{OSIRIS} is \qty{9}{\meter}, which is equal to a maximum water pressure of \qty{0.9}{\bar} the \ac{iPMT} has to withstand. Since this pressure is low and the \acp{PMT} are separated sufficiently, implosion protection is not necessary.

The electronics-stack is encapsulated inside a stainless steel shell. To ensure proper cooling of all electronics boards, this shell is filled with white oil. Via convection, the oil transports the heat from the electronics to the shell, where it gets transferred to the surrounding water. The water in the \ac{OSIRIS} tank is kept at a stable temperature of about \qty{21}{\celsius}.

A \ac{PMMA} ring  is glued to the neck of the \ac{PMT}. The electronics is mounted on the back of the \ac{PMT}, the shell is glued to the \ac{PMMA} ring and filled with oil (cf. to~\cite{PhDFengGao} for details of the potting). The oil expands under heating. If no countermeasures are taken, this expansion will damage the electronics. An air-filled \ac{HDPE} bottle is placed inside the shell to absorb the expansion of the oil.

Special measures have been taken to seal the joint between the cable and the shell. In the region, where the cable enters the shell, all wires are stripped down to the bare copper. For isolation, the cable is embedded in a low viscosity epoxy. It stops creeping of oil or water along the wires in both directions.

\begin{figure}[htb]
    \centering
    \includegraphics[width=0.5\textwidth]{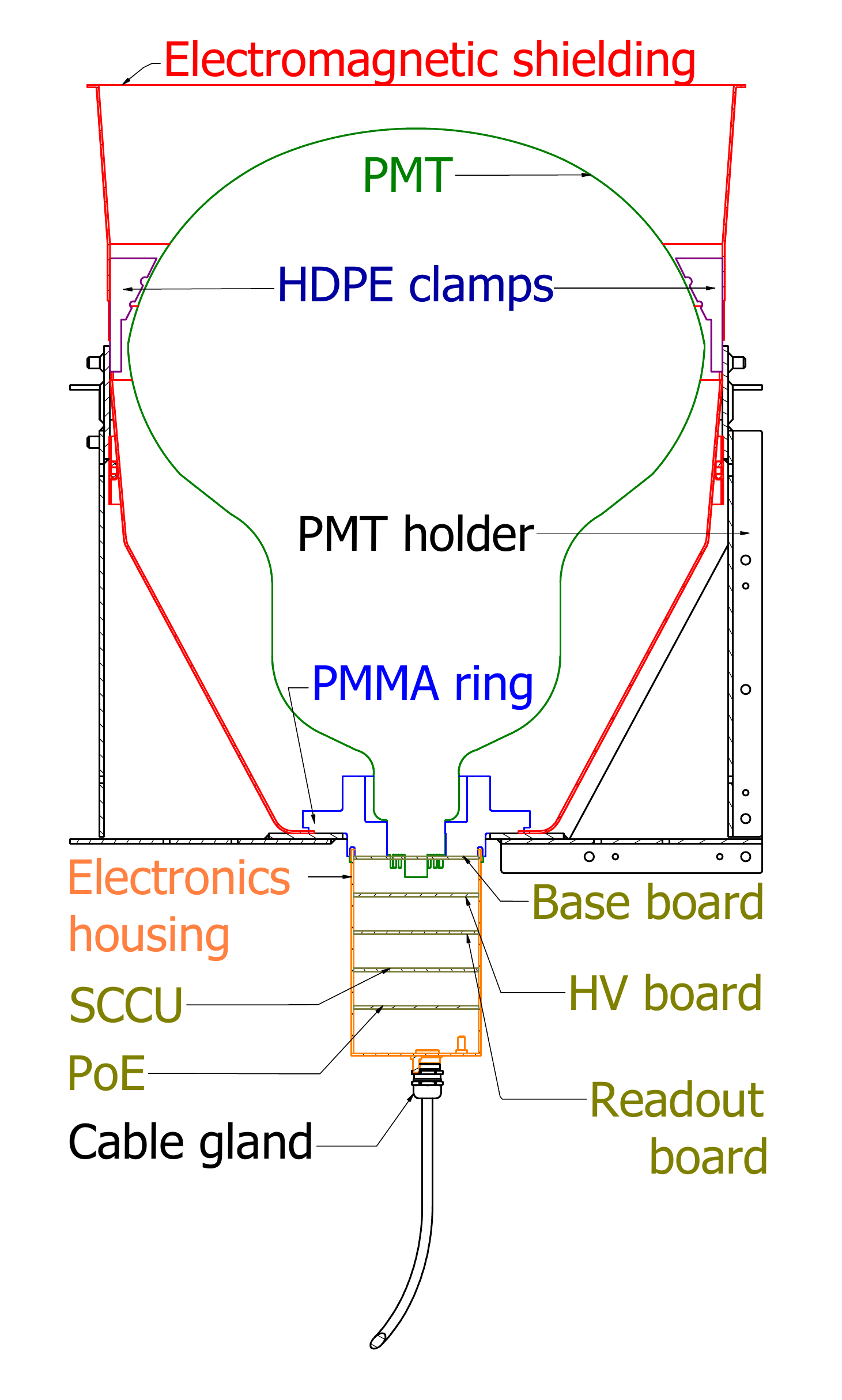}
    \caption{Overview of the \ac{iPMT} assembly.}
    \label{fig:iPMTassemblyOverview}
\end{figure}

The location of the items described above can be seen in Figure~\ref{fig:iPMTassemblyOverview}. The whole \ac{iPMT} is mounted together with its electromagnetic shield~\cite{MagneticShieldingPaper} in a stainless steel holder. This holder fixes the position and orientation of the \ac{iPMT} within \ac{OSIRIS}. For further support of the \ac{PMT}, four clamps, made from \ac{HDPE}, are located around the equator.

\section{\acl{SB}}
\label{sec:surface_board}
\begin{figure}[htbp]
    \centering
    \includegraphics[width=0.7\textwidth]{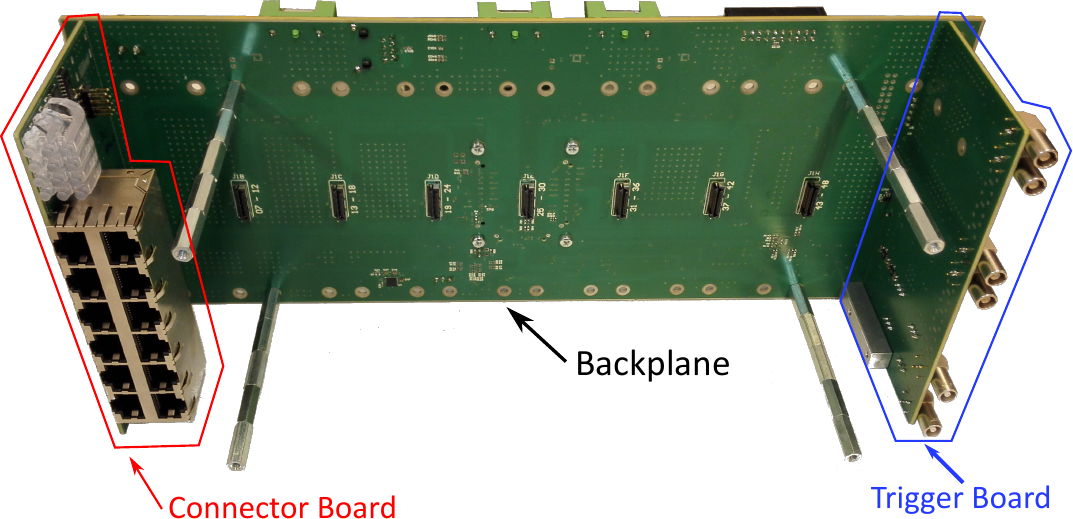}
    \caption{Image of the \ac{SB} with one Connector board and one Trigger Board.}
    \label{fig:sb_backplane}
\end{figure}

The main purpose of the \acf{SB} is the synchronisation of all \acp{iPMT}.
Each \ac{SB} hosts eight cards (Connector Boards) which can connect to six \acp{iPMT} each (cf. Figure~\ref{fig:sb_backplane}). In total, up to \qty{48}{\acp{iPMT}} can be connected to a single \ac{SB}. For \ac{OSIRIS}, two \acp{SB} are needed. One \ac{SB} acts as a \emph{master} and provides the \emph{slave} \ac{SB} with a clock and a datastream, which is forwarded without any further processing.

After the cables from the \acp{iPMT} arrive at the \ac{SB}, their Ethernet pairs are \remove{split} passively \add{separated}\R{A204} from the synchronous ones and further forwarded to a commercial Ethernet switch.
The synchronous connection to the \ac{iPMT} can be divided into a downlink (\ac{SB} to \ac{iPMT}) and an uplink with the reverse direction. All downlinks are fed with the same signal, which is generated by a single output of a \Xilinx~\ZYNQ on the \ac{SB}. This signal is distributed by clock splitters to the cable drivers of all \acp{iPMT}. For the uplinks, the signal of the \ac{iPMT} is routed from the cable receiver to a dedicated input for each \ac{iPMT}. This creates the possibility to measure the length of the cable \add{ by injecting a pulse on the downlink and measuring the travel time, when looped back inside the \ac{iPMT}}.\R{A210} The link \replace{quality}{integrity}\R{A211} can be measured using \ac{PRBS}. 

Programming \remove{of} the \ZYNQ and monitoring \remove{the} temperatures, voltages etc. on the \ac{SB}, is done by \replace{a}{one} \ac{SCCU} using \ac{JTAG} and \ac{I2C}. In addition, the \acp{UART} of the \ac{SCCU} are connected to the \ZYNQ for monitoring and debug purposes.

\subsection{LED \& Laser Trigger}
Each \ac{SB} provides two independent \add{periodic} trigger outputs via a Trigger Board. The trigger \replace{repetition frequency}{rate} can be adjusted in a range from \qty[exponent-to-prefix = true, scientific-notation=engineering]{0.12}{\hertz} to \qty[exponent-to-prefix = true, scientific-notation=engineering]{2.6e6}{\hertz}. The width of the output pulse can be configured from \qtyrange{16}{496}{\nano\second}. The outputs are used to trigger light emission from the Laser and LED calibration systems within \ac{OSIRIS}. A maximal rate of \qty{10}{\kilo\hertz} for the laser trigger is expected~\cite{OSIRISDesign}.
Each output pulse is tagged with a timestamp. This information is sent via Ethernet to the \ac{DAQ}, where this information is used to form calibration events.

\subsection{Synchronisation}
All \acp{iPMT} are supplied with the identical synchronous datastream with a bitrate of \qty{125}{\mega\bit\per\second}. A Manchester code is used on the stream which reduces the datarate to \qty{62.5}{\mega\bit\per\second}. A \qty{125}{\mega\hertz} clock is recovered from the datastream on the \ac{ROB} by the \ac{CDR}. The recovered clock is used as reference clock for VULCAN. By using the same frequency on all \acp{iPMT}, a frequency synchronisation is achieved (syntonization). Via the embedded datastream the \acp{iPMT} are partially synchronised. After this partial synchronisation, a time offset, which is dominated by the cable length difference between the \acp{iPMT}, remains. This offset can be determined by measurements of the electrical cable length or using pulsed Laser illumination events. A combination of both methods will provide a full synchronisation of all \acp{iPMT}.

\begin{figure}[htbp]
    \centering
    \includegraphics[trim=3.2cm 15.5cm 0.cm 6.4cm, width=0.7\textwidth, clip]{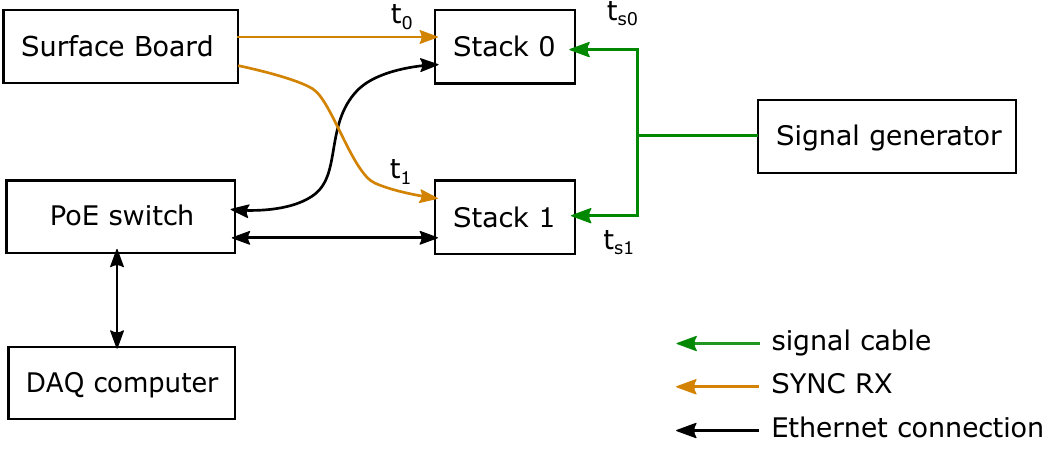}
    \caption{Sketch of the synchronisation setup. The different cable lengths to the electronics stacks (Stack 0 and Stack 1) are represented by the signal travel times $t_0$ and $t_1$.}
    \label{fig:sync_setup}
\end{figure}
\begin{figure}[htbp]
    \centering
    \subcaptionbox{\label{fig:sb_sync_single_pulse}}{\includegraphics[width=0.49\textwidth]{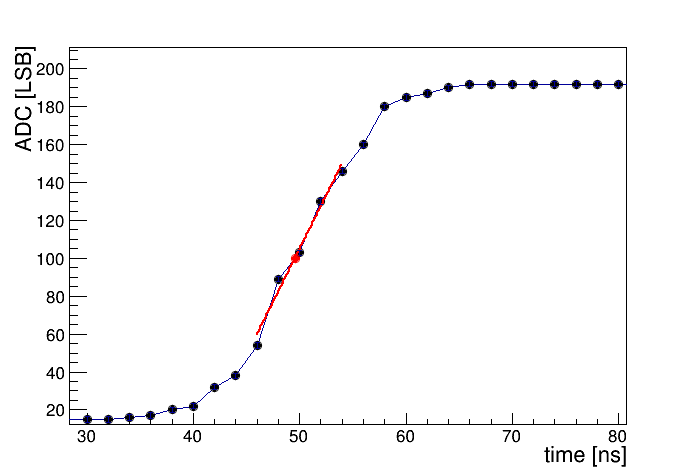}}
    \subcaptionbox{\label{fig:sb_time_diff_vs_time}}{\includegraphics[width=0.49\textwidth]{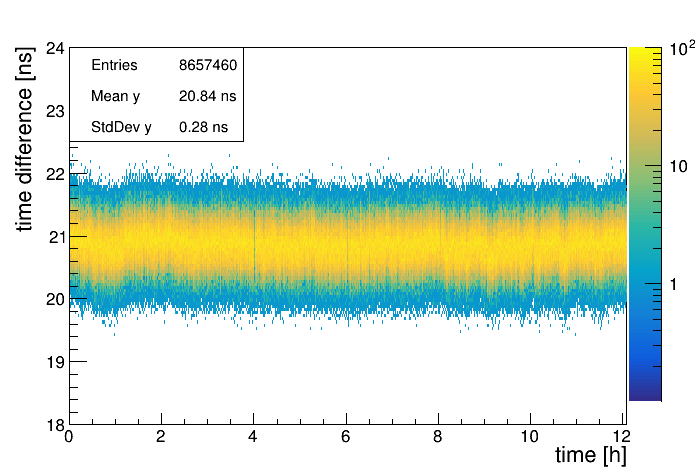}}
    \caption{\subref{fig:sb_sync_single_pulse} Zoom of a recorded waveform.
    \subref{fig:sb_time_diff_vs_time} Histogram of the synchronisation difference against the measurement time.}
\end{figure}

For testing the partial synchronisation, two half stacks (\ac{POE}-Board, \ac{SCCU} and \ac{ROB}) are connected to a pulse generator as a signal source. The cable length to both stacks is not matched (cf. Figure \ref{fig:sync_setup}). When the signal generator creates a pulse, this gets recorded by both stacks. Each stack tags this event with its own timestamp.
Figure \ref{fig:sb_sync_single_pulse} shows part of a single recorded waveform. To determine the proper timestamp, a linear function is fitted to the rising edge. The time when this interpolation crosses \qty{100}{\LSB} is used as the event time $t_{s0}$ and $t_{s1}$. Figure~\ref{fig:sb_time_diff_vs_time} shows the difference between both event times against the measurement time. The total measurement time shown in Figure~\ref{fig:sb_time_diff_vs_time} is divided into six times \qty{2}{\hour} due to file size constrains. The mean synchronisation difference between both used half-stacks is \qty{20.84 \pm 0.28}{\nano\second}.
The difference of \qty{20.84}{\nano\second} is determined by the different cable length. More relevant is the variation of the synchronisation over time, characterised by a standard deviation of \qty{0.28}{\nano\second}. It is well below the \ac{PMT}'s intrinsic \ac{TTS} of about \qty{1.1}{\nano\second}\footnote{Figure~\ref{fig:pmt_dist_tts} states a mean \ac{TTS} of \qty{2.6}{\nano\second} as \ac{FWHM}. With the assumption of a Gaussian distribution this number converts to a standard deviation of \qty{1.1}{\nano\second}.}.

\section{Slowcontrol of the \acs{iPMT} system}
\label{sec:slow_control}
In order to control all different parts of the \ac{iPMT}, an implementation based on \ac{EPICS}~\cite{EPICS} version {3.14.12.7} has been chosen.

For each subsystem, like \ac{I2C} or \ac{HV}, a dedicated \ac{IOC} handles the communication to the \ac{iPMT} using a custom implementation. Most of the endpoints are handled by the \ac{SCCU}. 
The conversion of the raw data into physical values is done by \ac{EPICS} conversion settings. 
The whole \ac{iPMT} can be monitored and controlled via \ac{EPICS}. It is possible to use a graphical interface as well as scripts to interact with the \acp{iPMT}.

The initial values, limits, and alarm settings of each record are derived from a \emph{MySQL} database. Each \ac{iPMT} can be configured individually. Most of the limits require tuning during the commissioning of \ac{OSIRIS}. Hence there is a default -- fail safe -- configuration in the database, which is used in case that no dedicated values are set for an \ac{iPMT}.

\section{Performance of a single \acs{iPMT}}
\label{sec:performance}
In this section the representative performance of a single \ac{iPMT} is shown.
For the following measurements, the \ac{iPMT} has been placed in a magnetic shielding box. An LED illuminates the photo cathode with short pulses of low intensity. LED and \ac{iPMT} share a common trigger. The response of the \ac{iPMT} to the light emission is recorded and analysed. For studies of the single photo electron signal, the light intensity of the LED has been tuned such\replace{,}{} that a pulse is observable in about \qty{10}{\%} of all recorded events. The synchronous detection of two or more photons is highly suppressed.

\begin{figure}[htbp]
    \centering
    \includegraphics[width=\textwidth]{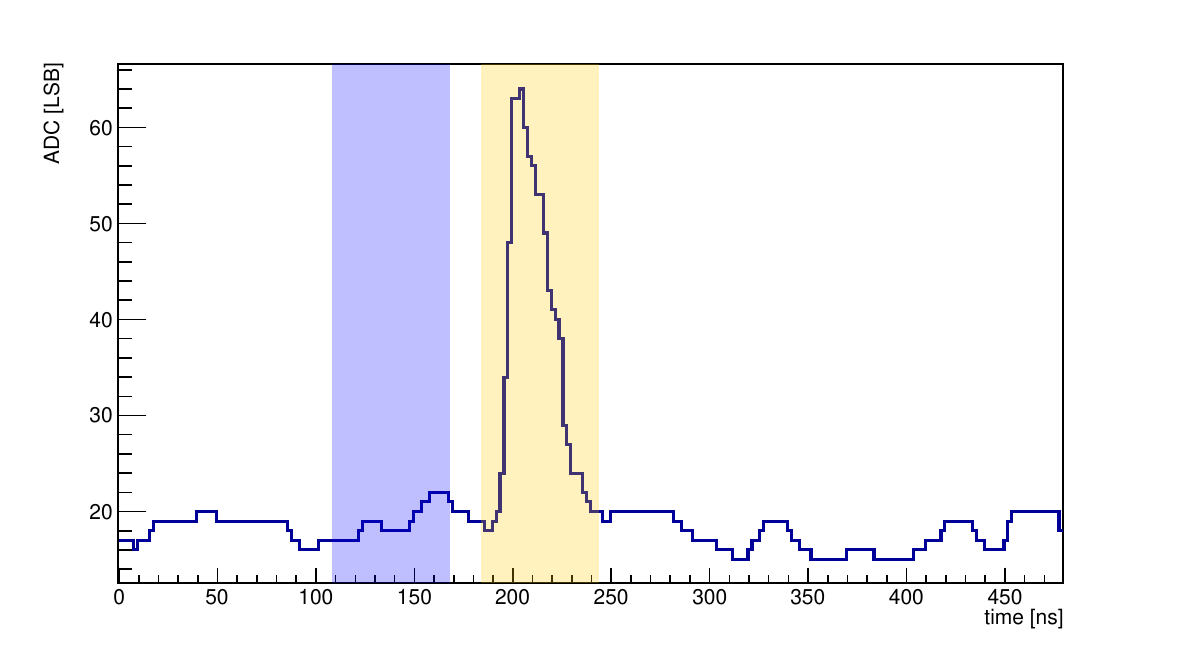}
    \caption{Exemplary waveform of a \ac{PE} with the baseline window marked in blue and the charge integration window marked in yellow.}
    \label{fig:exem_wv_pmt}
\end{figure}

Only those waveforms are processed whose average baseline in the baseline window (cf. Figure~\ref{fig:exem_wv_pmt}) -- from \qtyrange[evaluate-expression]{2*\baselineAVGstart}{2*\baselineAVGend}{\nano\second} -- is in a range of $\pm 1\sigma$ around the mean average baseline. The accepted average baseline values are \qtyrange{\CUTbaselineN}{\CUTbaselineP}{\LSB}.
%
%
By using this cut, the simple charge integration method works reliably and baseline fluctuations from earlier hits are rejected. 
Using a more advanced waveform reconstruction algorithm would supersede this cut.

\subsection{Charge characteristics}
\begin{figure}[htbp]
    \centering
    \includegraphics[width=\textwidth]{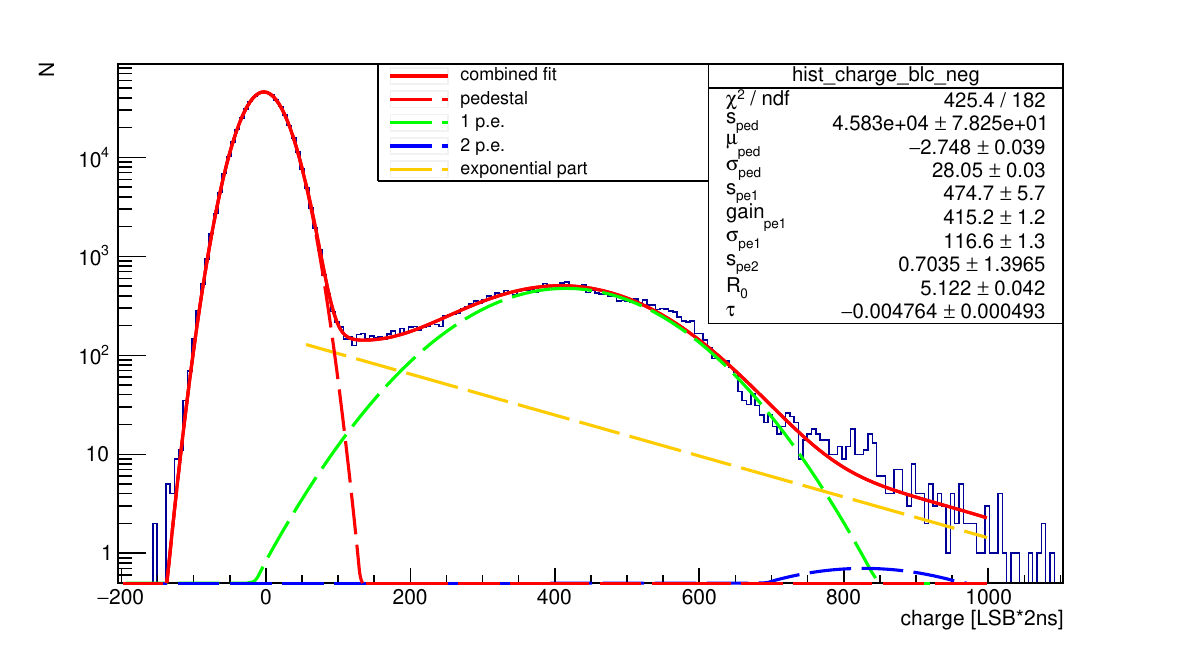}
    \caption{Single \acl{PE} charge spectrum.}
    \label{fig:exem_charge_spectrum}
\end{figure}

To analyse the charge, the waveform is corrected for the baseline, first. The average of the samples in the baseline window is subtracted from all samples. The charge integration window ranges from 
\qtyrange[evaluate-expression]{2*\IntegrationStart}{2*\IntegrationEnd}{\nano\second}
(cf. Figure~\ref{fig:exem_wv_pmt}). The histogram of these waveform integrals are shown in Figure~\ref{fig:exem_charge_spectrum}. It can be seen, that a tiny fraction of events with two \ac{PE} exists in the measurement as well.

The charge histogram is fitted with Equation~\ref{eq:ptv_fit_func} to extract the gain and other parameters.
\begin{align}
\label{eq:ptv_fit_func}
f_{\text{fit}}(x) &= f_{\text{ped}}(x)+f_{\text{pe1}}(x)+f_{\text{pe2}}(x)+f_{\text{exp}}(x)\\
%
f_{\text{ped}}(x) &= s_{\text{ped}}\cdot e^{-0.5\cdot ((x-\mu_{\text{ped}})/\sigma_{\text{ped}})^2}\\
f_{\text{pe1}}(x) &= s_{\text{pe1}}\cdot e^{-0.5\cdot ((x-(\mu_{\text{ped}}+\text{gain}_{\text{pe1}}))/\sigma_{\text{pe1}})^2}\\
f_{\text{pe2}}(x) &= s_{\text{pe2}}\cdot e^{-0.5\cdot ((x-(\mu_{\text{ped}}+2\cdot \text{gain}_{\text{pe1}}))/(\sqrt{2}\sigma_{\text{pe1}}))^2}\\
f_{\text{exp}}(x) &= 
\begin{dcases*}
e^{R_{0}+\tau \cdot (x-x_v)} & if $x > x_v$\\
0 & \textrm{otherwise}
\end{dcases*}.
\end{align}
$f_{\text{ped}}(x)$ is a Gaussian component to model the noise, also denoted as pedestal or noise peak. 
$f_{\text{pe1}}(x)$ and $f_{\text{pe2}}(x)$ are Gaussian functions that describe the charge signals corresponding to one, respective two \aclp{PE}.
Since the \ac{PMT} is a linear amplifier for this low number of \ac{PE}, the respective Gaussian mean values are correlated via the variables $\text{gain}_{\text{pe1}}$ and $\sigma_{\text{pe1}}$.
$f_{\text{exp}}(x)$ is an empirical approach to include other effects into the model to better describe especially the region between pedestal peak and single \ac{PE} peak.
The variable $x_v$ is the starting point of this fit on the falling edge of the pedestal peak~\cite{WysotzkiPhD,DC_pmt_test_2011}.
The \ac{SNR} is given by
\begin{equation}
\text{SNR}_\text{charge} = \frac{ \text{gain}_{\text{pe1}} }{ \sigma_{\text{ped}} } = \add{\num[round-mode=uncertainty]{\chargeSNR +- \chargeSNRUncert}}\, .
\end{equation}
The noise -- in terms of charge -- is $\frac{1}{\text{SNR}} = \qty[evaluate-expression, round-mode=places,round-precision=2]{\chargeNoiseWidth}{\ac{PE}}$ wide. The charge resolution is calculated as
\begin{equation}
\text{Resolution}_\text{charge} = \frac{ \sigma_{\text{pe1}}} { \text{gain}_{\text{pe1}} } = \add{\qty[round-mode=uncertainty]{\chargeResolution +- \chargeResolutionUncert}{\%}}\, .
\end{equation}
The charge resolution is not fully determined by the readout electronics. The amplification process inside the \ac{PMT} limits  this value~\cite{BscKuhlbusch}.

From the fit (Equation \ref{eq:ptv_fit_func}), the peak to valley ratio can be determined. It is calculated from the height of the single \ac{PE} peak divided by the number of entries in the valley between the single \ac{PE} peak and the noise peak. For this measurement the value is \num{\PeakToValley}, compared to the value of \num{3.08} measured by the vendor.

\subsection{Characteristics of the waveform maximum}
\begin{figure}[htbp]
    \centering
    \includegraphics[width=\textwidth]{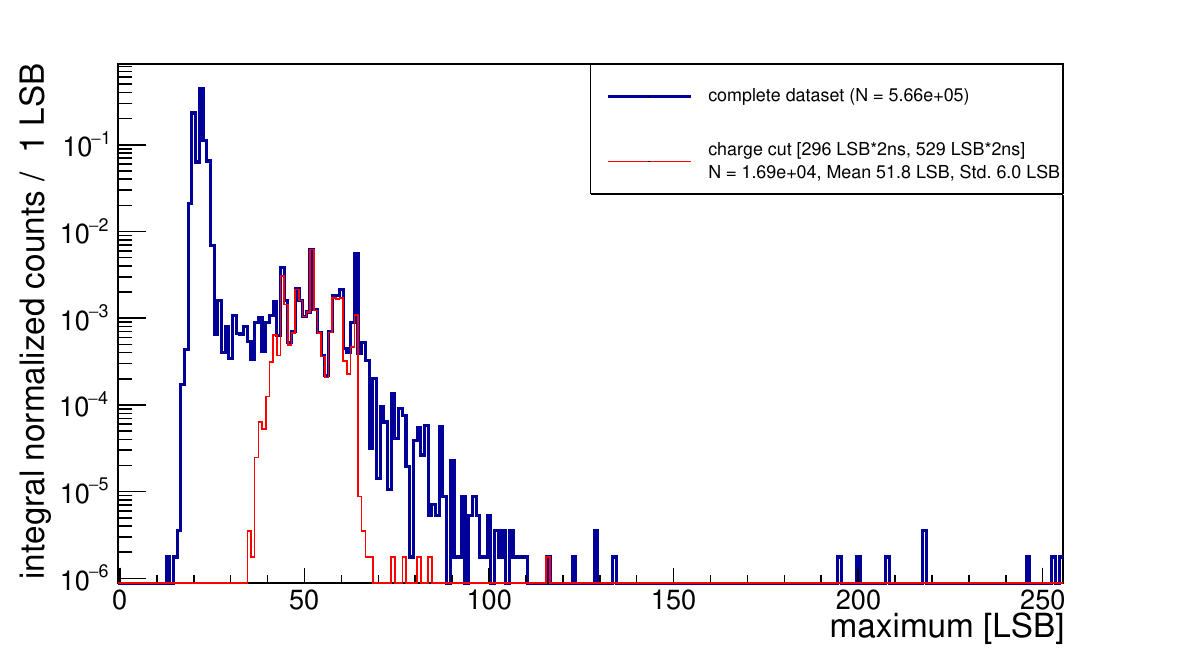}
    \caption{Maximum distribution in charge integration window.}
    \label{fig:amp_dist}
\end{figure}

This study is done using the raw \ac{ADC} waveforms without any additional calibration. A distribution of the maximum \ac{ADC} value within the integration window is shown in Figure~\ref{fig:amp_dist}. Those waveforms which have a charge of $\pm 1\sigma$ around the single \ac{PE} peak (see Figure~\ref{fig:exem_charge_spectrum}) are plotted in red. From this distribution it can be seen that the  single \ac{PE} maximum is on average \qty{ \amplSPEmean +- \amplSPEsigma}{\LSB}. With the baseline of \qty[round-mode=uncertainty]{ \amplBASELINEmean +- \amplBASELINEsigma }{\LSB}, 
the baseline corrected maximum of a single \ac{PE} is \qty[round-mode=uncertainty,round-precision=3]{ \amplCalcSPEheight +- \amplCalcSPEheightSigma }{\LSB}. Using these values, the dynamic range of the high gain receiver with a full scale value ${\text{FS}_\text{HG} = \qty{255}{\LSB}}$ can be calculated as
\begin{equation}
    \text{N}_\text{PE, HG} = \frac{ \text{FS}_\text{HG} - \text{baseline}_\text{HG} }{\text{maximum}_\text{SPE, HG} - \text{baseline}_\text{HG} } 
    = \qty[round-mode=uncertainty,round-precision=3]{ \NPEinHG +- \NPEinHGsigma }{\ac{PE}}
\end{equation}
%
As for the charge, the \ac{SNR} can be calculated accordingly to
\begin{equation}
\text{SNR}_\text{maximum} = \frac{ \text{maximum}_\text{SPE, HG} - \text{baseline}_\text{HG} }{ \sigma_{\text{baseline}} } 
= \num[evaluate-expression, round-mode=places,round-precision=2]{ (\amplSPEmean - \amplBASELINEmean) / \amplBASELINEsigma }\, .
\end{equation}
This value is comparable to the charge based value. The resolution can also be derived from the measurement shown in Figure~\ref{fig:amp_dist} as
\begin{equation}
\text{Resolution}_\text{maximum} = \frac{ \sigma_{\text{ampl}}} { \text{maximum}_\text{SPE, HG} - \text{baseline}_\text{HG} } = \qty[evaluate-expression, round-mode=places,round-precision=2]{100 * ( \amplSPEsigma / (\amplSPEmean - \amplBASELINEmean) )}{\%}\, .
\end{equation}
The resolution of the waveform maximum is less than the charge resolution. But for the maximum only the highest value inside the integration window is used. For the charge, all information from the whole pulse is used. Using a more dedicated waveform reconstruction algorithm might increase the maximum resolution as well.

\subsection{Self triggered charge spectrum}
In \ac{OSIRIS}, the \acp{iPMT} operate in self-triggered mode. They send out a waveform with a timestamp once their local trigger condition is met. With this acquisition mode, a charge spectrum can be recorded as well. Figure~\ref{fig:self_trg_charge_spectrum} depicts multiple charge spectra recorded with increasing trigger thresholds. It starts at the lowest trigger threshold, which can be recorded without data loss. The charge spectrum obtained with the external trigger is plotted for comparison, too.
\begin{figure}[htbp]
    \centering
    \includegraphics[width=\textwidth]{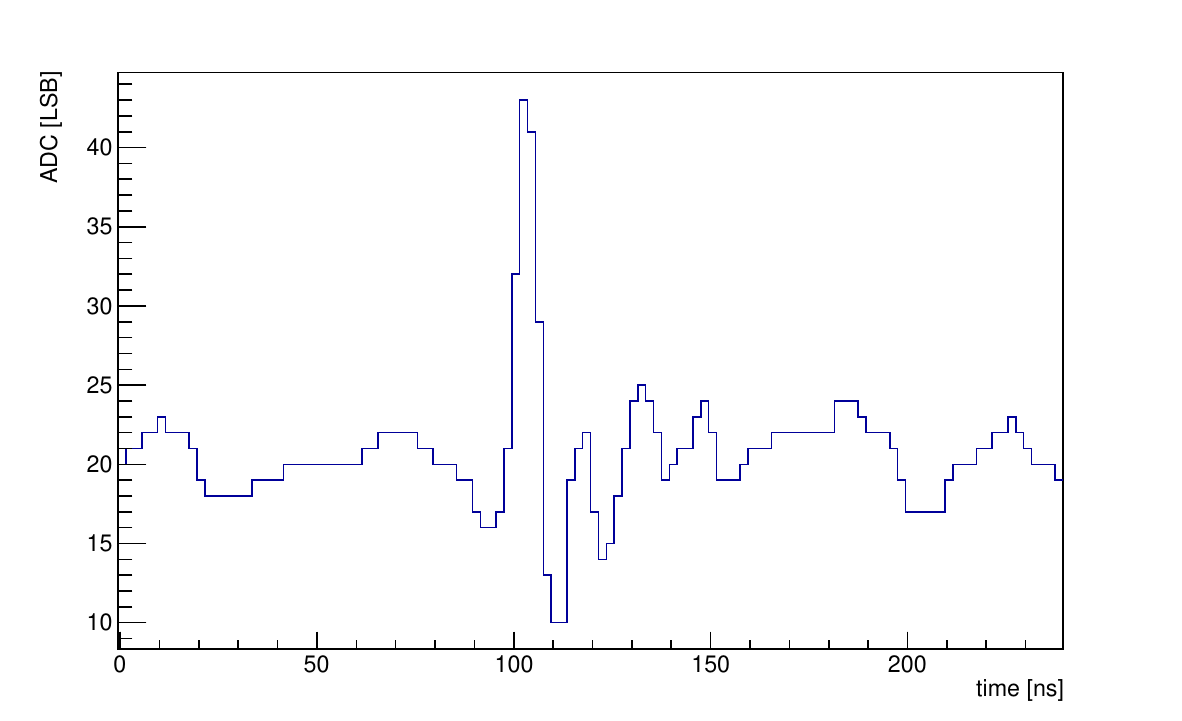}
    \caption{Example of a waveform which contributes to the noise in the self-triggered charge spectrum.}
    \label{fig:self_trg_noisy_wv}
\end{figure}
 The trigger thresholds are given in raw \ac{ADC} counts. After subtraction of the baseline~(\qty{\amplBASELINEmean}{\LSB}), the minimal pulse height can be derived, which is accepted with this setting.
\begin{figure}[htbp]
    \centering
    \includegraphics[width=\textwidth]{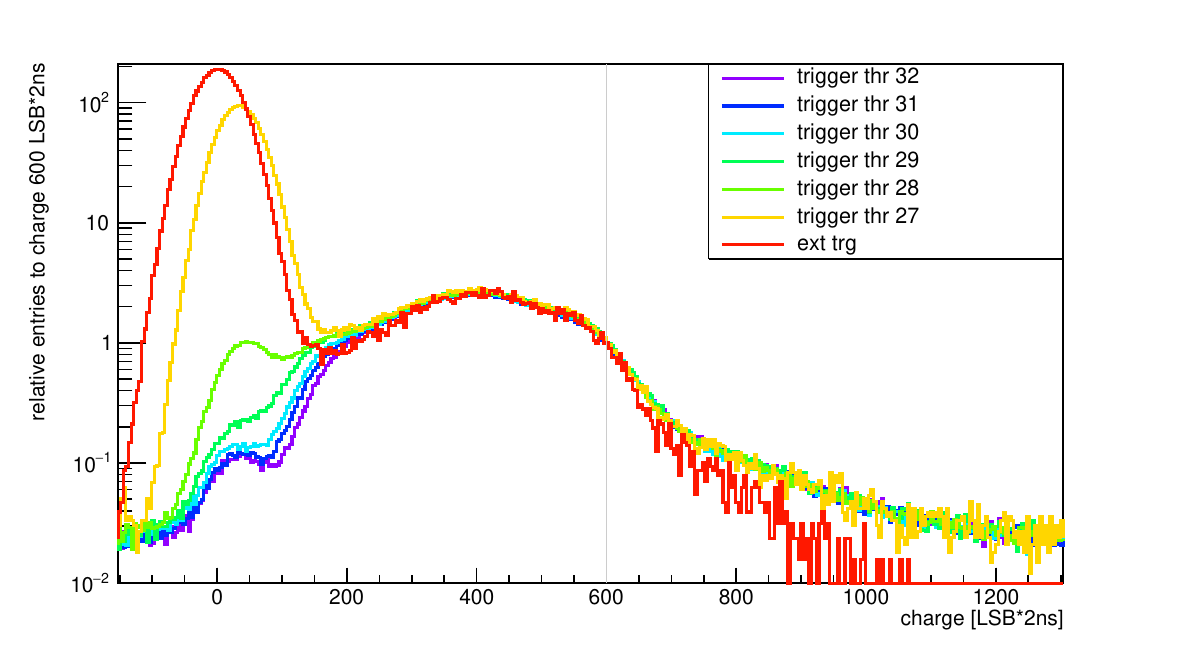}
    \caption{Charge spectra externally triggered and self-triggered. All different spectra are normalised to  \qty{600}{\LSB \cdot 2\nano\second}. }
    \label{fig:self_trg_charge_spectrum}
\end{figure}
Figure~\ref{fig:self_trg_charge_spectrum} shows that the noise suppression increases with the trigger threshold. However, there are still few noise events remaining, which cause a peak between \qtyrange{0}{100}{\LSB \cdot 2\nano\second}. These events have a damped oscillation signature of unknown source as shown in Figure~\ref{fig:self_trg_noisy_wv}.
With increasing trigger threshold, the detection threshold in terms of light intensity increases as well.
\FloatBarrier
\section{Summary / Conclusions}

The \ac{iPMT} concept is a novel \ac{PMT} readout concept based on the close integration of the \ac{PMT} and electronics. We presented a detailed description of the system components. It is shown that a partial synchronisation can be achieved, which has a sufficiently low uncertainty.
The performance of one exemplary \ac{iPMT} is presented, focusing on low light level characteristics.
\add{In the charge performance this \ac{iPMT} achieved a signal to noise ratio of \num[round-mode=uncertainty]{\chargeSNR +- \chargeSNRUncert} and a resolution of \qty[round-mode=uncertainty]{\chargeResolution +- \chargeResolutionUncert}{\%} for single photon events}\R{A322_concl}.

For \ac{OSIRIS}, \qty{75}{\acp{iPMT}} have been successfully produced, tested and shipped to the \ac{JUNO} site for installation. \remove{Due to unforeseen problems and the necessary repairs of the {iPMT}, they are not installed in the first phase of {OSIRIS}. They will be used in the future upgrade of {OSIRIS} focusing on the search for solar $pp$ neutrinos.}\R{A322}

\acknowledgments
This work was supported by the Deutsche Forschungsgemeinschaft (DFG), the Helmholtz Association, and the Cluster of Excellence PRISMA+ in Germany. \R{A325} 


\bibliographystyle{JHEP}
\bibliography{literature}

\ifistoreview\newpage
\section*{Answers to review comments}
\textbf{Comments from a A. Triossi}
{\parindent0pt

L.8: "it is not evident to me why it should be valid expecially for neutrino experiments"\\
\textbf{Answer:} We reformulated this sentence to keep the focus on neutrino experiments, which have a huge volume but only few sensors. \lr{A8} You are absolutely right, the cable length for other detectors increase as well, but for example for CMS at LHC, the signals are digitised as early as possible and sent to the DAQ via fiber. 

\vspace{1em}

L. 68: afford -> effort\\
\textbf{Answer:} Fixed

\vspace{1em}

L. 73: "the digital signal degradates as well as the analog, but the analog shape is not anymore carrying the information. Nevertheless a degradation of the rising/falling edge (slope) of the signal limits bandwidth and transmission rate."\\
\textbf{Answer:} This is why we have included, that the length of the cable is limited as well. 

\vspace{1em}

L. 80: "please specify the mechanism for discrimination: leading edge, constant fraction..."\\
\textbf{Answer:}
added "by e.g. a simple threshold discriminator or more advanced pulse detection techniques" \lr{A80}. It is not necessarily a simple discriminator since more advanced pulse detection circuits can be implemented in the FPGA as well.

\vspace{1em}

L. 83: "What is "concept"? Electronics, detector, block diagram, etc..."  \\
\textbf{Answer:} added "readout" to specify which concept was reused.\lr{A83}

\vspace{1em}

L. 83: "It is not explained why this detector electronics it is not used in the central detector and it is not clear to me what do you mean with "iteration". I would expand this sentence giving more explanations or remove it at all. "\\
\textbf{Answer:} I'm not sure, whether there is a citeable document, giving the explanation, why the BX concept has not made it into JUNO. \\
We would like to mention it here, because the readout concept has not been developed by us, but we have iterated it once again in order to tune / adjust some parameters of the hardware design to our needs and use VULCAN as PMT frontend + digitizer.

\vspace{1em}

L. 90: "It is not clear if the R15343 is used in OSIRIS or not. Did the problem on the voltage unit force to change PMT?"\\
\textbf{Answer:} Exactly, we were forced to change the PMTs from iPMTs to LPMTs for OSIRIS. We removed the sentence from \lr{A90} and \lr{A322} and inserted it into the introduction \lr{A322_fix}.

\vspace{1em}

L. 123: "Of what?" -> added "of the internal clocks" \lr{A123} 

\vspace{1em}

L. 124: "It would be nice to write here why 500Mbps is enough for OSIRIS"\\
\textbf{Answer:} We removed the part that explains that VULCAN can also run with 1GSps since this information is not relevant for the paper.

\vspace{1em}

L. 131: "It is not clear to me what you mean with concatenated"\\
\textbf{Answer:} We removed this part of the sentence. This avoids confusion of the reader and since this part of information is not used any more in the paper, we can remove it without loosing information.

\vspace{1em}

L. 132: "maybe the logic? it should not be a real processor" \lr{A132}\\
\textbf{Answer:} Inside VULCAN it is called data processor \cite[p. 45]{MuralidharanPhd}. We clarified this by using the term "data processing logic".

\vspace{1em}

L. 139: "a" -> included in \lr{A139}

\vspace{1em}

L. 142: "the SCCU itself" -> included in \lr{A142}

\vspace{1em}

L. 145: "It is a modification of UDP or it is a custom protocol over UDP?"\\
\textbf{Answer:} Good point, we clarified this in the text \lr{A145}

\vspace{1em}

L. 165: "You could add that this syncronous link will be presented later in the paper"\\
\textbf{Answer:} The synchronous link has been already mentioned in the ROB section.

\vspace{1em}

L. 167: "a" -> fixed

\vspace{1em}

L. 204: "separated"\\
\textbf{Answer:} We reformulated this to "passively separated" \lr{A204}. For us it is important to emphasize that the splitting / separation is done in a passive way.

\vspace{1em}

L. 210: "...by injecting a pulse on the dowlink and measuring the time of flight when loopback to the uplink in the iPMT?"\\
\textbf{Answer:} We added this explanation how to measure the cable length.\lr{A210}

\vspace{1em}

L. 211: "integrity?" -> replaced \lr{A211}

\vspace{1em}

L. 212: remove "of" -> done

\vspace{1em}

L. 212: remove "the" -> done

\vspace{1em}

L. 213: "one" -> replaced "a" by "one"

\vspace{1em}

L. 216: "periodic trigger" -> added periodic

\vspace{1em}

L. 216: "Rate" -> replaced

\vspace{1em}

L. 258: "A" \\
\textbf{Answer:} "An LED" should be correct since "LED" starts with a vowel sound

\vspace{1em}

L. 261: "no comma" -> removed

\vspace{1em}

L. 278: "Please, cite also "Qualification Tests of 474 Photomultiplier Tubes for the Inner Detector of the Double Chooz Experiment" where this technique is used as well"\\
\textbf{Answer:} Added a reference to the paper.

\vspace{1em}

L. 307: "Figure 14 and Figure 13 should be swapped"\\
\textbf{Answer:} The reason to have a nice take home message, is, why we have swapped Figure 14 and Figure 13. So we end with a nice spectrum and not with the waveform, which shows a noise event. 
Doubling an image from section 7 is what we wanted to avoid, because the paper is not too long.

\vspace{1em}

L. 322: "In my opinion, I would remove this part from the conclusion. Maybe adding one of the nice figure of merit found in section 7, it would be a nice take home message at the end of the paper \lr{A322}"\\
\textbf{Answer:} We removed the part from the summary and added it in the introduction \lr{A322_fix}. In the conclusion we added a part stating the performance of the system, see \lr{A322_concl}.

\vspace{1em}

L. 325: "Probably this should go in a dedicated section
Forn JINST author's maunal:
Acknowledgments. The command \verb|\acknowledgments| starts a new non-numbered section
where the acknowledgments can be placed. It usually resides before the bibliography, or at the end of the introduction."\\
\textbf{Answer:} Thank you for the advice. We included the \verb|\acknowledgments| section \lr{A325}

}\fi

\end{document}